\documentclass[%
 aip,
 amsmath,amssymb,
 reprint,%
floatfix,
]{revtex4-1}

\usepackage{graphicx}
\usepackage{dcolumn}
\usepackage{bm}

\usepackage[utf8]{inputenc}
\usepackage[T1]{fontenc}
\usepackage{mathptmx}
\usepackage{etoolbox}
\usepackage{enumitem}
\usepackage{siunitx}
\usepackage{hyperref}

\makeatletter
\def\@email#1#2{%
 \endgroup
 \patchcmd{\titleblock@produce}
  {\frontmatter@RRAPformat}
  {\frontmatter@RRAPformat{\produce@RRAP{*#1\href{mailto:#2}{#2}}}\frontmatter@RRAPformat}
  {}{}
}%
\makeatother
\begin{document}

\preprint{AIP/123-QED}

\title[An ultra-low field SQUID magnetometer]{An ultra-low field SQUID magnetometer for measuring antiferromagnetic and weakly remanent magnetic materials at low temperatures}
\author{Michael Paulsen}
\thanks{The following article has been submitted to the Review of Scientific Instruments. After it is published, it will be found at \href{https://publishing.aip.org/resources/librarians/products/journals/}{Link}.}
\email[Corresponding author, e-mail: ]{michael.paulsen@ptb.de}
\affiliation{Physikalisch-Technische Bundesanstalt Berlin (PTB), 7.6 Cryosensors, Abbestrasse 2-12, 10587 Berlin, Germany}

\author{Julian Lindner}
\affiliation{Helmholtz-Zentrum Berlin f\"{u}r Materialien und Energie (HZB), Sample Environment Group,  Hahn-Meitner-Platz 1, 14109 Berlin, Germany}

\author{Bastian Klemke}
\affiliation{Helmholtz-Zentrum Berlin f\"{u}r Materialien und Energie (HZB), Sample Environment Group,  Hahn-Meitner-Platz 1, 14109 Berlin, Germany}

\author{J\"{o}rn Beyer}
\affiliation{Physikalisch-Technische Bundesanstalt Berlin (PTB), 7.6 Cryosensors, Abbestrasse 2-12, 10587 Berlin, Germany}

\author{Michael Fechner}
\affiliation{Max Planck Institute for the Structure and Dynamics of Matter, Luruper Chaussee 149, 22761 Hamburg, Germany}
\author{\\Dennis Meier}
\affiliation{Norwegian University of Science and Technology (NTNU), Department of Materials Science and Engineering, Sem S{\ae}landsvei 12, N-7034 Trondheim, Norway}
\affiliation{Center for Quantum Spintronics, Department of Physics, NTNU, Trondheim 7491, Norway}
\author{Klaus Kiefer}
\affiliation{Helmholtz-Zentrum Berlin f\"{u}r Materialien und Energie (HZB), Sample Environment Group,  Hahn-Meitner-Platz 1, 14109 Berlin, Germany}
\date{\today}

\begin{abstract}
A novel setup for the measurement of magnetic fields external to certain antiferromagnets and generally weakly remanent magnetic materials is presented. The setup features a highly sensitive Super Conducting Quantum Interference Device (SQUID) magnetometer with a magnetic field resolution $\sim$\,\SI{10}{\femto\tesla}, non-electric thermalization of the sample space for a temperature range of 1.5\,--\,\SI{65}{\kelvin} with a non-electric sample movement drive and optical position encoding. To minimize magnetic susceptibility effects, the setup components are degaussed and realized with plastic materials in sample proximity. Running the setup in magnetically shielded rooms allows for a well-defined ultra low magnetic background field well below \SI{150}{\nano\tesla} in situ. The setup enables studies of inherently weak magnetic materials which cannot be measured with high field susceptibility setups, optical methods or neutron scattering techniques, giving new opportunities for the research on e.g. spin-spiral multiferroics, skyrmion materials and spin ices.
\end{abstract}

\maketitle



\section{Introduction}
\label{sec:intro}
\begin{figure}[h!]
\centering
\includegraphics[width=0.99\linewidth]{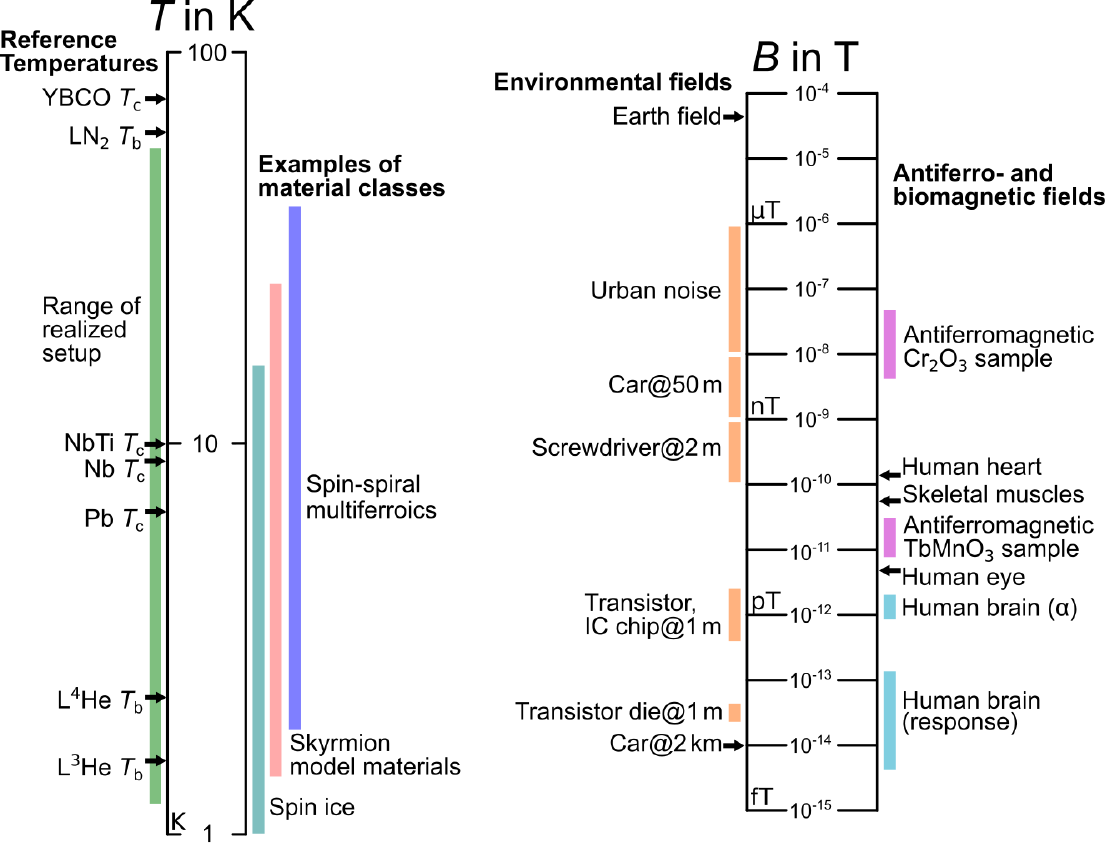}
\caption{Overview of temperatures \cite{pobell2007matter} $T_{x}$ ($x$\,=\,b\,[boiling],\,c\,[critical]), materials and magnetic field strengths \cite{vrba2002magnetoencephalography} that are relevant for the type of measurements addressed by the developed setup. Significant examples of material systems are spin-spiral multiferroics such as orthorhombic rare-earth manganites \cite{goto2004ferroelectricity,kimura2007spiral} \textit{R}MnO$_{3}$ (\textit{R}\,=\,Gd, Tb, Dy), manganese tungstate \cite{arkenbout2006ferroelectricity} and the olivine \cite{white2012coupling} Mn$_{2}$GeO$_{4}$, classical skyrmion materials \cite{muehlbauer2009skyrmion,munzer2010skyrmion} (e.g. MnSi, Fe$_{1-x}$Co$_{x}$Si) and spin ice materials like the pyrochlore \cite{matsuhira2001novel} Dy$_{2}$Ti$_{2}$O$_{7}$.}
\label{fig:scope_of_fields_temperatures}
\end{figure}

The external magnetic field of a material is given by its susceptibility as a function of the magnetic background field and, depending on the micromagnetic moments of the sample, its inherent magnetism. Conventional ferro-/ferrimagnets, such as compounds of iron, cobalt and nickel, exhibit pronounced magnetic fields $>$\,\SI{}{\micro\tesla} that can readily be measured by inductive coils or fluxgate devices. In contrast, the complex magnetic order of spin ice, spin-spiral multiferroics, skyrmions lattices, antiferromagnets, and other emergent functional materials is often hard to detect with characteristic magnetic field values as low as \SI{}{\pico\tesla}. Furthermore, depending on the material's susceptibility, it can be difficult to ensure non-invasive magnetization measurements and to probe the intrinsic magnetic properties free from interfering magnetic background contributions. Achieving such experimental conditions is challenging due to the earth magnetic field, remanence of setup materials and currents of electronic devices, as depicted in Fig. \ref{fig:scope_of_fields_temperatures}.
\newline \indent The motivation for realizing an appropriate ultra low-field variable temperature magnetometer were seminal magnetization measurements on antiferromagnets. Generally, antiferromagnets have been studied for several decades in fundamental research and, more recently, as materials of interest in spintronic devices \cite{Jungwirth2016}. As there is no net external dipole field in the antiferromagnetic phase, measuring this state usually involves sophisticated methods e.g. neutron scattering facilities \cite{brockhouse1953antiferromagnetic, kenzelmann2005magnetic} or susceptibility setups involving high magnetic fields \cite{buchner2018tutorial}. An instructive classical example of an antiferromagnetic solid state system with an ultra-weak external magnetic field is Cr$_2$O$_3$. It was predicted \cite{dzyaloshinskii1992external} to exhibit very weak, higher order net external magnetization ($\sim$\,\SI{10}{\nano\tesla} for a single domain, spherical sample with radius $\sim$\,\SI{5}{\milli\meter}). The few confirmed measurements of the respective quadrupolar magnetic fields were all conducted using dedicated SQUID setups which are not commercially available \cite{astrov1994quadrupole, astrov1996external}. Common to these setups is that the SQUID sensor and its superconducting pick-up coil need to be operated at an ideally constant low temperature ($<$\,\SI{9}{\kelvin} [\SI{77}{\kelvin}] for low [high] temperature superconductors). Also, the pick-up coil needs to be close ($\leq$\,\SI{15}{\milli\meter}) to the sample to cause a sufficiently strong magnetic signal. Since SQUID magnetometers detect magnetic flux \textit{change}, the sample is typically moved relative to the pick-up coil. Furthermore, the magnetic background field must be kept extremely low to avoid magnetization of the sample due to its magnetic susceptibility. Notably, the magnetic shielding was not quantitatively described in the constant- and variable-temperature setups in the literature \cite{astrov1994quadrupole, astrov1996external, borovik-romanov97}. This is a crucial issue as unwanted magnetization may lead to false quadrupole-like signals in case of anisotropic magnetic susceptibility. It should be noted that commercially available SQUID measurement systems typically involve electrical heaters and large magnets\cite{buchner2018tutorial}. While these can nominally be degaussed down to zero field, it is not clear if the actual background field within such a setup is significantly below the necessary 1\,Oe, i.e. \SI{100}{\micro\tesla} to avoid false signals due to anisotropic susceptibility\cite{astrov1996external}. Furthermore, the gradiometric pick-up coil of commercial systems is usually optimized for measuring the magnetic dipole moment of the sample, rather than higher order magnetic contributions.\\
\indent The realization of a magnetometer with very low magnetic background and high magnetic field resolution was recently described \cite{SQUID_HZB_non_VTI} by the authors. In that setup, the sample was kept at constant liquid helium temperature ($T_{\text{LHe}}$=\SI{4.2}{\kelvin}) and slowly rotated, using a plastic gravity-driven pendulum motor along with an optical encoder, $\sim$\,\SI{4}{\milli\meter} in front of a superconducting pick-up coil connected to a SQUID sensor. The magnetometer can advantageously be used in combination with a Glass-Fiber-reinforced Cryostat (GFC) inside of a magnetically shielded environment such as the Berlin Magnetically Shielded Room-2 (BMSR-2) \cite{knappe2008influence} or a smaller ultra-low magnetic field shielding\cite{sun2020limits}.\\
\indent In this work, we present the realization of a novel experimental setup \cite{SQUID_HZB_VTMI}, which enables temperature-dependent measurements of external fields of magnetically ordered materials. The sample temperature is varied without electric heating and the setup consists of very low remanence materials, such as PolyEther Ether Ketone (PEEK) and PolyVinyl Chloride (PVC).

\section{Description of Experimental Setup}
\begin{figure}[h]
\centering
\includegraphics[width=0.72\linewidth]{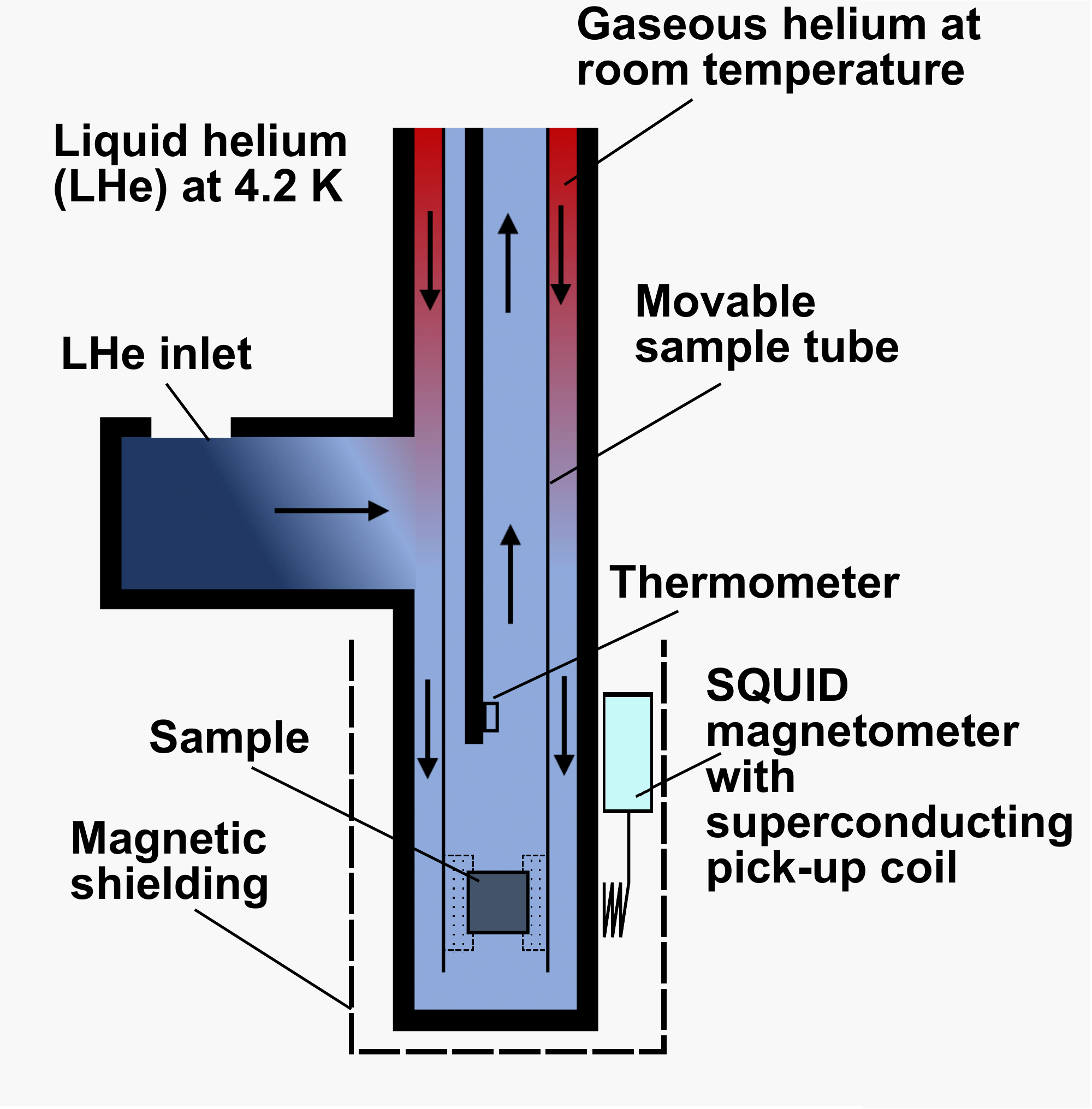}
\caption{Schematic measurement principle of the VTMI.}
\label{fig:Measurement_principle}
\end{figure}

\begin{figure}[ht]
\centering
\includegraphics[width=0.99\linewidth]{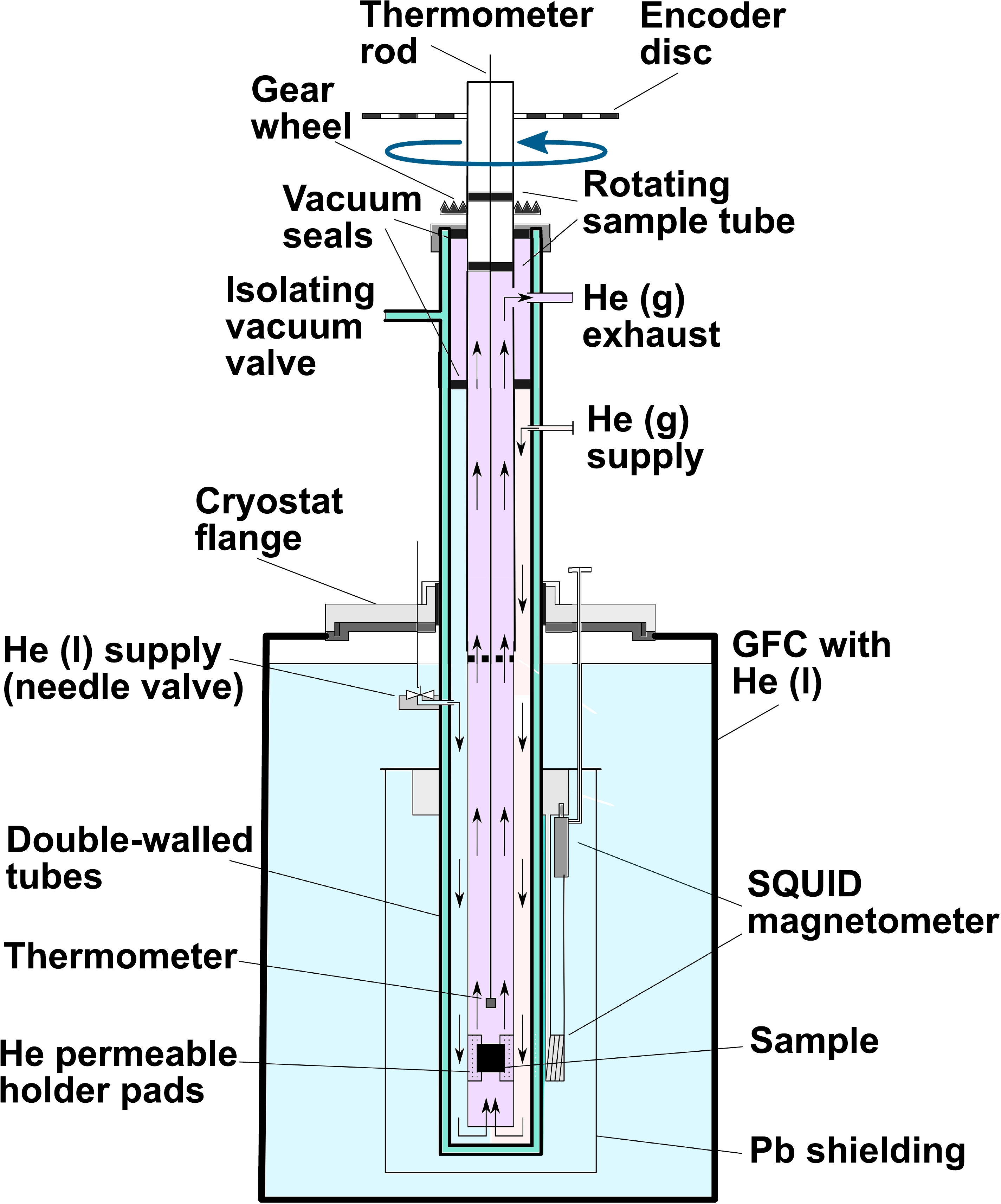}
\caption{Schematic view of the realized VTMI\cite{SQUID_HZB_VTMI} placed in a GFC.}
\label{fig:Sketch_Flow_and_Colour}
\end{figure}

\begin{figure}[ht]
\centering
\includegraphics[width=0.72\linewidth]{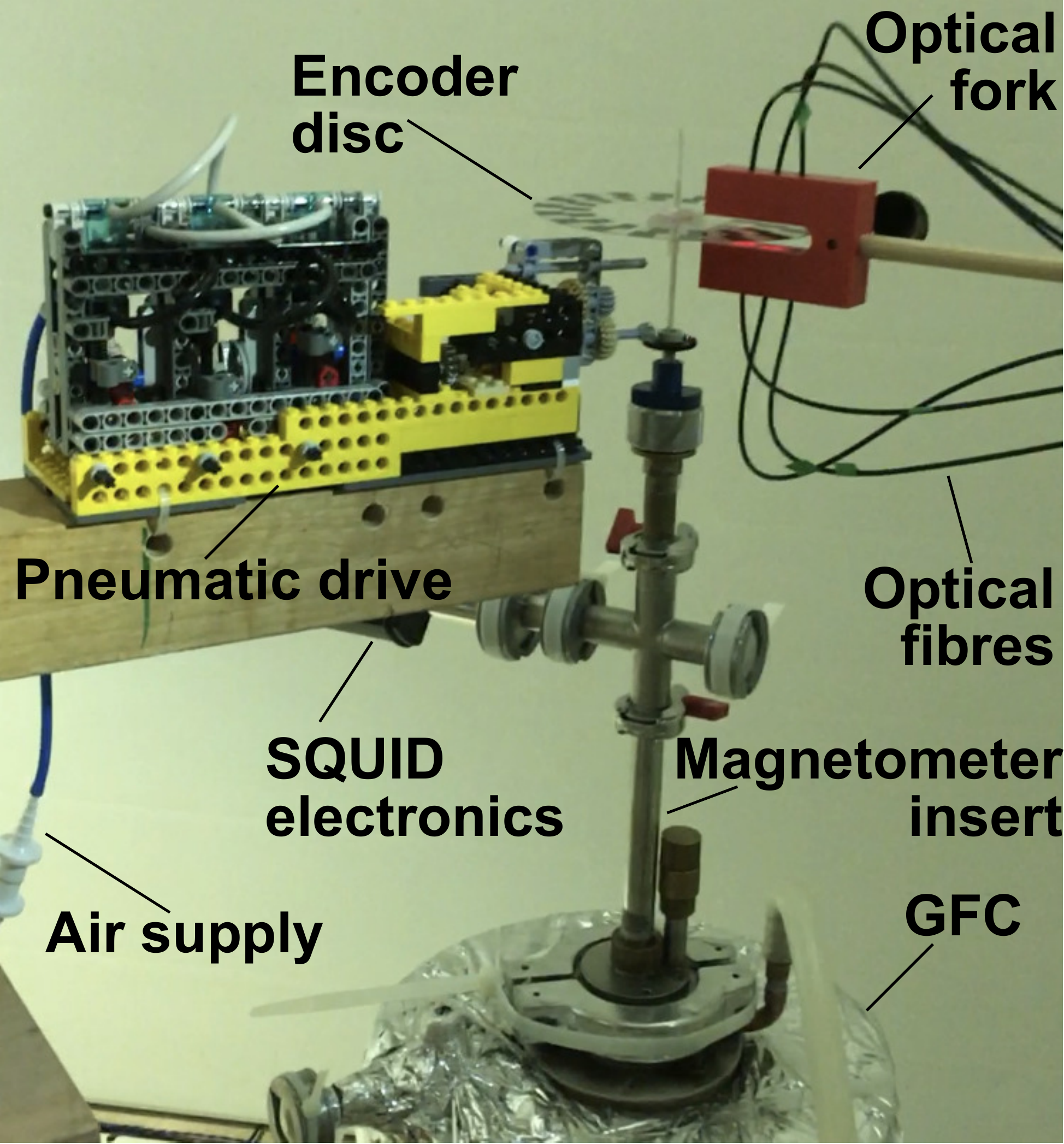}
\caption{The non-electric sample movement drive and optical position encoder in operation with a liquid helium temperature magnetometer insert\cite{SQUID_HZB_non_VTI}.}
\label{fig:Pneumatic_drive}
\end{figure}

The measurement principle is based on varying the temperature of the sample while moving/rotating it in front of a magnetometer in an ultra-low magnetic field. This is realized with a custom-built continuous Variable Temperature Magnetometer Insert (VTMI), as schematically depicted in Fig. \ref{fig:Measurement_principle}.
Liquid and gaseous helium enter and then flow downwards into the VTMI. At the bottom, the mixture enters the sample tube where it flows upwards and thermalizes the sample. A small thermometer mounted to the end of a small hollow temperature holder is placed centrally above the sample to measure the temperature of the upwardly flowing helium mixture. The sample tube can be moved to rotate the sample in front of a SQUID magnetometer. In order to perform such measurements in practice, the VTMI fulfills additional requirements.
\begin{itemize}
  \item The SQUID magnetometer is based on the superconductivity of niobium (Nb) and is operated at liquid helium temperature. A superconducting pick-up coil consisting of niobium titanium (NbTi) is placed in close proximity to the sample while an isolation vacuum surrounding the sample chamber is realized which requires a further layer of material. Thus, an additional short transfer tube to bring the liquid helium into the sample chamber is needed.
  \item Only low remanence magnetic materials should be used for the construction, especially regarding moving parts and all parts near the sample and SQUID to avoid magnetic background.
  \item To enable rotation and vertical movement of the sample holder tube, a rotation-compatible solution for extracting the helium mixture past the sample is called for, along with seals, which enable these movements while remaining gas-tight. Further tight seals are needed to ensure that the thermometer tube remains fixed during the measurements.
\end{itemize}	
The design of the VTMI is depicted in Fig. \ref{fig:Sketch_Flow_and_Colour} and was realized as follows. Double-walled tubes, made of PEEK in its lower part and stainless steel in its top part, are connected to a PEEK needle valve housing. This design ensures liquid helium supply via the GFC, while a gas valve controls the supply of exterior compressed helium gas at room temperature. The double-walled tubes have a valve that is connected to a pump ensuring an isolating vacuum.
\newline \indent A carbon fiber sample holder tube is kept in position via several vacuum seals. Its bottom part connects to a light plastic tube, e.g. a plastic straw, in which the sample is fixed with helium-permeable holding pads. The top part features two small holes and is located in ambient room temperature, where it connects to an exhaust valve at its top part, such that the liquid-helium mixture can flow past the sample. A hollow rod within the sample holder tube holds a small Cernox\textregistered\,\,thermometer (CX-1050-BC-HT) in the gas flow centrally above the sample. Several vacuum seals ensure that the sample holder tube can rotate while the thermometer rod remains fixed and centered using a centering piece such that helium gas can flow towards the exhaust valve. The thermometer wires run through the hollow thermometer rod to electronics at room temperature.
\newline \noindent The sample movement is provided by an ultra-low field drive, which is realized as a three cylinder pneumatic engine using commercially available LEGO\textregistered \,\,and custom-built plastic parts, as depicted in Fig. \ref{fig:Pneumatic_drive}. While the pneumatic engine itself runs at about 120\,rpm at a pressure of $\approx$\,\SI{1.5}{\bar} it is geared down several times to increase the torque which is needed for rotating the sample holder tube between the seals. The resulting rotation speed of the sample holder is $\approx$\,0.5\,rpm. The sample position is determined using a custom-built optical system, which is realized as a three-channel quadrature encoder consisting of SensoPart FL 70 light sensors, optical fiber cables along with an optical fork barrier and a plastic encoder disc.
\newline \noindent A PVC mounting supports the SQUID magnetometer and an optional superconducting lead (Pb) shielding. These components and the needle valve piece of the double-walled tubes are completely immersed in the liquid helium of the GFC and attached to the bottom part of the double-walled tube. The SQUID magnetometer is based on a PTB C6XXL1 single-stage current sensor \cite{Dru07}. Due to the Nb materials used in the chip, the sensor is operated at the very stable temperature of LHe at \SI{4.2}{\kelvin}. The control and readout of the sensor was done using Magnicon GmbH XXF-1 electronics. The output voltage of the SQUID electronics was read out with a Keithley 2010 multimeter. The current sensor is connected to a superconducting pick-up coil, consisting of NbTi-wire with a diameter of \SI{0.102}{\milli\meter} around a hollow PEEK tube. It has an outer diameter of \SI{10.5}{\milli\meter} and a coil of eleven turns with an inductance $\approx$\,\SI{900}{\nano\henry} was realized around it. The leads to the coil were carefully twisted to avoid parasitic inductance, which was estimated to be approximately \SI{50}{\nano\henry} as the leads had an approximate length of \SI{0.1}{\meter}. The complete measurement setup consists of electronics and vacuum pumps outside of a magnetic shielding, which contains a GFC with the custom-built VTMI.

\section{Operating Principle}
Before operation, the sample is placed within the sample tube in front of the SQUID magnetometer. While still at room temperature, an isolating vacuum of $\sim10^{-4}$\,mbar is pumped. The sample holder tube is carefully flooded with gaseous helium at a slight overpressure of around \SI{10}{\milli\bar}. The VTMI is cooled down by careful lowering into the helium-filled GFC and by fixing it at the cryostat flange such that the liquid helium level is well above the needle valve housing. The needle valve is opened to allow liquid helium to enter from the GFC into the sample chamber. To set the temperature, room-tempered helium gas is lead into the sample chamber and when the exhaust valve is kept open, the helium mixture flows through the outer sample holder tube and the sample is thermalized. To achieve a low consumption of helium and to avoid turbulence, which could cause a jiggle of the sample, the setup is operated at minimal flows. For temperatures above the boiling temperature of liquid helium at normal pressure, i.e. $>$\,\SI{4.2}{\kelvin}, gas flows corresponding to between 10 and \SI{25}{\milli\bar} of overpressure in the exhaust pipe are used.
\newline \indent During measurement operation, the temperature must be set manually by adjusting the helium gas valve and liquid helium needle valve, respectively. If the temperature at the sample location is intended to decline, the needle valve may be opened a little further or one may gradually close the helium gas valve. To increase the temperature, one proceeds vice versa. The settings of the valves should result in a pressure of the mixture flow within the verified range corresponding to 10\,--\,\SI{25}{\milli\bar}. Comparably high temperatures  $>$\SI{100}{\kelvin} can necessitate the complete closing of the needle valve. To reach temperatures below the boiling point of liquid helium at normal pressure, the setup makes use of the decreasing boiling temperature at diminishing pressure. Therefore, the sample chamber must be filled with some liquid helium and then the needle valve must be closed further, partially or completely. That way, sample temperatures as low as \SI{1.5}{\kelvin} can be reached.\\
\indent During development, tests of the variable temperature insert under realistic ambient conditions, with a sample dummy and two thermometers (one attached to the sample dummy and one at standard position), were performed. The temperature difference was lower than \SI{1}{K}. The in situ temperature range proved to be 1.5\,--\,65$\,$K with a temperature stability $\pm2$\,K of several minutes.
The helium consumption was approximately 1.5\,l/h, which allows for over 150\,rotations of the sample within one run, corresponding to approximately four hours of measurement time for the 8\,l GFC that we used.
When the realized setup was placed in the magnetically shielded room BMSR-2 at the PTB Berlin, the magnetic background field in situ was measured to be significantly less than \SI{150}{\nano\tesla} using hand-held fluxgate devices. The main magnetic contribution came from a remanent steel part in the needle valve, which in principle could be replaced with a PEEK version to allow for $<$\,\SI{10}{\nano\tesla} in situ magnetic background.
\section{Prototypical Measurement}
\begin{figure}[ht]
\centering
\includegraphics[width=0.99\linewidth]{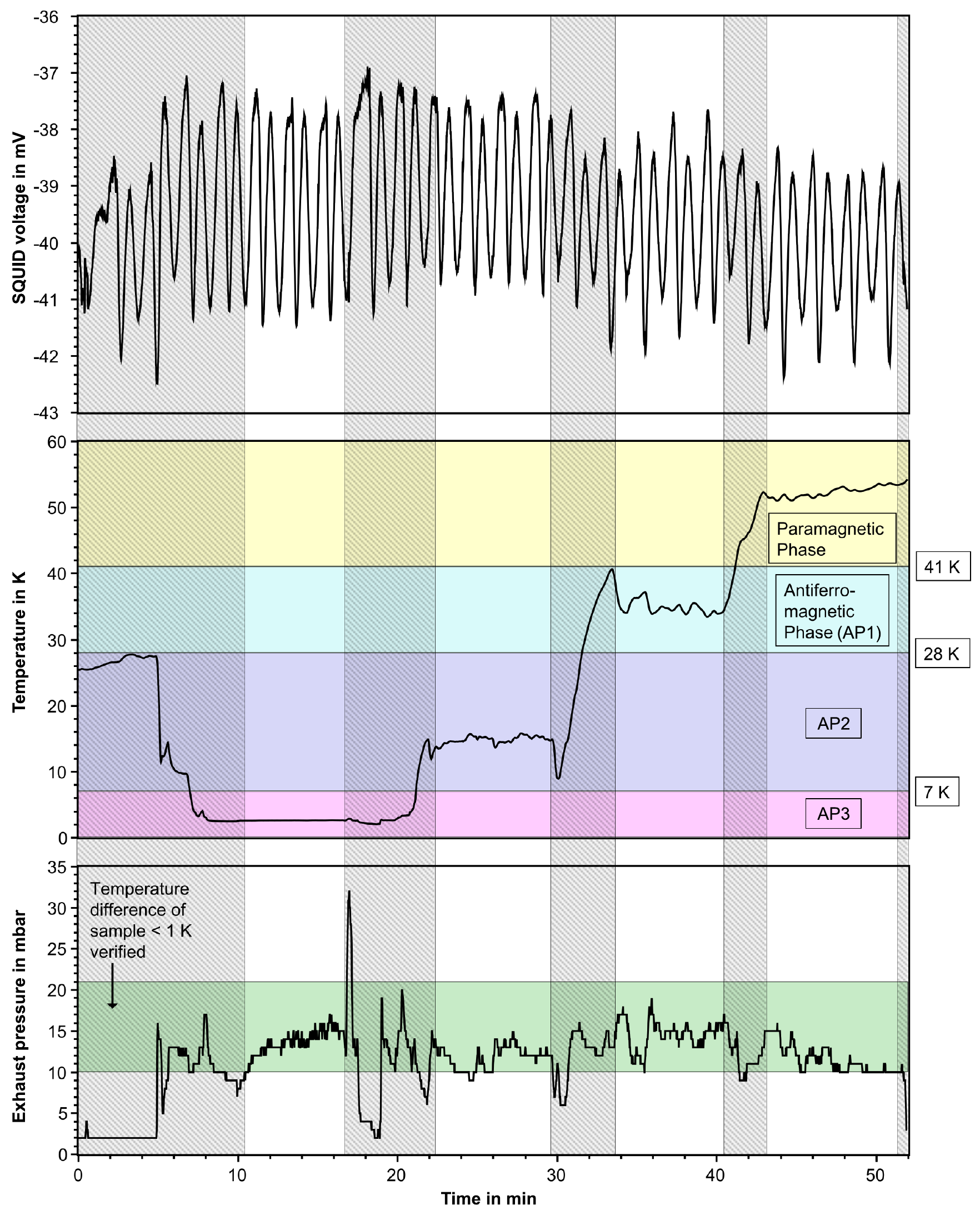}
\caption{Raw data plot of a test measurement using the novel setup: a TbMnO$_3$ sample was rotated at $\approx$\,0.5\,rpm. The output SQUID voltage [\textbf{top}], thermometer temperature [\textbf{middle}] and exhaust pressure [\textbf{bottom}] are shown, with transitions and stable regions depicted as \textbf{shaded} and \textbf{transparent}, respectively.}
\label{fig:Example_measurement_with_setup}
\end{figure}

In order to demonstrate the sensitivity of our setup, we performed temperature-dependent measurements on the spin-spiral multiferroic TbMnO$_{3}$ \cite{kimura2003magnetic}. The material exhibits different magnetic phase transitions at cryogenic temperature, going from paramagnetic (PM) to sinusoidal (AP1) to cycloidal (AP2) to Tb$^{3+}$ induced (AP3) antiferromagnetic order as explained elsewhere \cite{kenzelmann2005magnetic,kimura2007spiral}. Most important for this work, the magnetic order and transition temperatures are well-known and it is established that TbMnO$_{3}$ displays a compensated spin structure that allows only for weak higher-order contributions, whereas magnetic dipole contributions are forbidden by symmetry, with the temperature intervals of interest\cite{kimura2007spiral} $\text{PM}>T_{\text{N}}=41$\,K $>$ AP1 $>T_{\text{C}}=28$\,K $>$ AP2 $>T_{\text{A}}=7\,$K $>$ AP3.\\
\indent Fig. \ref{fig:Example_measurement_with_setup} presents a temperature-dependent measurement gained on a TbMnO$_{3}$ single crystal with side length $\approx$\,\SI{5}{\milli\meter}. The recorded output SQUID voltage, thermometer temperature and the exhaust pressure are plotted over the measurement time, which was 52\,minutes. At the start, while rotating the sample using the pneumatic drive, the liquid helium needle valve was opened and the valve for room-temperature helium gas was gradually closed. This caused the temperature to reach a stable \SI{1.5}{\kelvin}, i.e. the AP3 state of the sample, for six minutes. By closing the needle valve and opening the helium gas valve appropriately, the AP2, AP1 and finally the PM state of the sample were measured. Thus, the intended functionality of the developed setup was successfully verified.

\section{Summary}
A novel setup designed for measuring ultra-small magnetic fields was presented. It is readily applicable to measure weak magnetic remanence of a wide range of solid state materials including spin-spiral multiferroics\cite{vopson2015fundamentals}, spin ices\cite{gingras2011spin} and 3D printed nanomaterials \cite{lowa20193d} which are used as phantoms in magnetic resonance imaging research\cite{arenas2022}. Compared to other setups reported in literature and commercial SQUID systems, the apparatus features several advantages. The sample movement and thermalization are entirely composed of non-metallic and non-electric components, which greatly reduces magnetic interactions with the sample along with customized state-of-the-art SQUID magnetometry. The setup performed excellently in a controlled, magnetically shielded room. The realized prototype can measure the external magnetic field of small (diameters $\sim$\,1--\SI{7}{\milli\meter}) very weakly remanent ($\sim$\,\SI{0.1}{\pico\tesla} at sample surface) samples in the temperature range 1.5\,--\,\SI{65}{\kelvin} with an in-situ background field well below \SI{150}{\nano\tesla}, giving new opportunities for the study of complex low-remanent materials with otherwise hard-to-measure magnetic properties.

%
%

%

\begin{acknowledgments}
The technical support of Martin Petsche at HZB, Lars Schikowski at PTB, Monique Klemm and Sassan AliValiollahi at Magnicon GmbH is gratefully acknowledged. The setup was tested at the BMSR-2 within the Deutsche Forschungsgemeinschaft core facility \textit{Metrology for Ultra-Low Magnetic Fields} with the generous help of Rainer Körber, Allard Schnabel and Jens Voigt. The authors thank Tsuyoshi Kimura for providing the TbMnO$_{3}$ single crystal for our experiments. We further thank Niclas Scholz (manufacturing), Dirk Gutkelch (development) and Norbert Löwa (research) of the B-smart Lab at PTB for the support with additive manufacturing concerning a 3D-printed magnetic nanomaterial phantom sample that was kindly provided to us. DM acknowledges funding from the Research Council of Norway, project number 263228 and support through its Centres of Excellence funding scheme, Project No. 262633, “QuSpin”, and thanks NTNU for support via the Onsager Fellowship Program and the Outstanding Academic Fellows Program.
\end{acknowledgments}

\bibliography{SQUID_setup}

\end{document}